# Inverse Temperature Dependence of Nuclear Quantum Effects in DNA Base Pairs

Wei Fang,[†] Ji Chen,[‡] Mariana Rossi,[¶] Yexin Feng,[§] Xin-Zheng Li,[*,||] and Angelos Michaelides[*,‡]

[†]*Thomas Young Centre, London Centre for Nanotechnology and Department of Chemistry, University College London, London WC1E 6BT, UK*

[‡]*Thomas Young Centre, London Centre for Nanotechnology and Department of Physics and Astronomy, University College London, London WC1E 6BT, UK*

[¶]*Physical and Theoretical Chemistry Lab, University of Oxford, South Parks Road, OX1 3QZ Oxford, UK*

[§]*School of Physics and Electronics, Hunan University, Changsha 410082, P. R. China*

[||]*International Center for Quantum Materials, School of Physics and Collaborative Innovation Center of Quantum Matter, Peking University, Beijing 100871, P. R. China*

E-mail: xzli@pku.edu.cn; angelos.michaelides@ucl.ac.uk

**Abstract**

Despite the inherently quantum mechanical nature of hydrogen bonding it is unclear how nuclear quantum effects (NQEs) alter the strengths of hydrogen bonds. With this in mind we use *ab initio* path integral molecular dynamics to determine the absolute contribution of NQEs to the binding in DNA base pair complexes; arguably the most important hydrogen bonded systems of all. We find that depending on temperature NQEs can either strengthen or weaken the binding within the hydrogen bonded complexes. As a somewhat counter-intuitive consequence, NQEs can have a smaller impact on hydrogen bond strengths at cryogenic temperatures than at room temperature. We rationalize this in terms of a competition of NQEs between low frequency and high frequency vibrational modes. Extending this idea, we also propose a simple model to predict the temperature dependence of NQEs on hydrogen bond strengths in general.

# Graphical TOC Entry

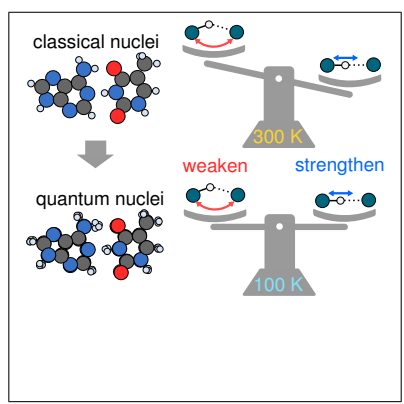

It has been said that without hydrogen bonds (HBs) all wooden structures would collapse, cement would crumble, oceans would vaporize, and all living things would disintegrate into inanimate matter.[1] Whilst the concept of the HB dates back to at least the 1920s[2] and it is now well-defined,[3] the small mass of the proton means that HBs and intrinsically quantum mechanical and zero point energy and tunneling can be of critical importance. Quantum fluctuations involving HBs are crucial, for example, in biological processes such as DNA tautomerization[4–6] and enzyme reactions.[7–12] It is also known that hydrogendated and deuterated chemicals can have different biochemical potencies; a fact that is now enthusiastically being exploited within the pharmaceutical industry through the development of novel classes of deuterated drugs. Nonetheless fundamental understanding of the quantum nature of HBs is far from complete, with a general understanding of how and to what extent nuclear quantum effects (NQEs) influence the strengths of HBs yet to be established. Given that strength is arguably the most important property of any bond this seems to represent a fairly significant gap in understanding.

Indirect information about the role of NQEs on hydrogen bond strengths can be made through isotopic substitution experiments. These have shown that upon replacing hydrogen with deuterium the lengths of HBs can change; a so-called secondary geometric isotope effect.[13,14] HBs can get shorter or longer depending on the material and the extent of the change can vary from one material to the next and with temperature (see e.g.[15–18]). Interestingly, very small changes in structure can lead to large variations in physical properties. For example, the lattice constants of the hydrogenated and deuterated versions of the ferroelectric material squaric acid differ by 1% yet their ferroelectric to paraelectric transition temperatures differ by *ca.* 200 K.[16,17] Secondary geometric isotope effects such as these have been explained with the help of theory and simulation,[18–23] and notably it has been argued that the direction and extent of the change upon isotopic substitution depends on the strength of the HB.[18] This, in turn, has been rationalized with a theory of competing quantum effects where it is said that quantum delocalization along the HB helps to shorten the bond but delocalization out of the plane acts to lengthen it.[19,22,24,25] Indeed this concept has proved to be useful in explaining a host of phenomena in e.g. liquid water, ice and biomolecules[25–31] and has seen recent experimental verification.[32] Nonetheless, most work to date has focused on geometrical properties and direct information on how and to what extent NQEs influence the strengths of HBs[33] is desirable. Although the total contribution of NQEs to HB strength is likely to be small, small energies are often important. This is particularly true in biology where structures and processes are governed by a delicate balance of interactions.[31] The cost to unzip double stranded DNA in solution is, for example, only *ca.* 20 - 100 meV (*ca.* 0.5 - 2 kcal/mol, or 1 - 4 $k_{\mathrm{B}}T$) per base pair duplex at room temperature.[34–36] Similarly the melting temperature of DNA strands is so exquisitely sensitive that substituting just one out of a thousand base pairs leads to a measurable change in melting temperature.[37]

In this work we use computer simulations to directly evaluate the influence of NQEs on the binding free energy of DNA base pair complexes. These hydrogen bonded complexes are not only crucial to life, but have also gained significant interest in nanotechnology.[38] We specifically examined Watson-Crick hydrogen bonded base pair complexes of adenine-thymine (AT) and cytosine-guanine (CG) in the gas phase. The focus is on understanding how quantum effects alter the duplex hydrogen binding interaction in the dimers; this is, of course, an integral interaction to DNA binding and is, for example, the key parameter in nearest neighbor[36] and coarse-grained models[39] of DNA binding. Although stacking interactions and solvent effects are relevant to the unzipping and melting of real double stranded DNA, by focusing exclusively on the duplex hydrogen bonding interaction we are able to unambiguously understand the role played by NQEs. The particular computational techniques employed involve density functional theory (DFT) for a description of the potential energy surface in conjunction with path integral molecular dynamics (PIMD), which together enables equilibrium thermal properties including NQEs to be rigorously accounted for. This methodology has been widely applied to tackle a host of chemical problems in the gas and condensed phases (see e.g. ref. 17,18,29–31,33,40–46). Furthermore by combining

PIMD with thermodynamic integration we can explicitly calculate the binding free energy change upon transforming the system from one containing classical nuclei to one containing quantum nuclei. With this approach we find, at room temperature, that NQEs increase the interaction strength in both complexes by $ca.$ 0.5 kcal/mol or $ca.$ 1 $k_{\mathrm{B}}T$. At 100 K, a temperature appropriate for preserving DNA information and a temperature at which NQEs are generally expected to be more significant, we find that the impact of NQEs on the binding energy is smaller. Analysis reveals that this surprising temperature dependence arises from a competition of quantum effects associated with low and high frequency vibrational modes. Extending our findings from DNA base pairs, we use our physical understanding of competing quantum nuclear effects to propose a simple model to estimate the temperature dependence of NQEs on binding free energies of hydrogen bonded complexes in general.

Our PIMD simulations were performed with the CP2K[47,48] code connected to the i-PI wrapper.[49] A full account of the computational set-up is given in the supporting information (SI) section II and here we outline the key features. The PIGLET thermostat[50] was used for an efficient sampling of the imaginary-time path integrals. At 300 (100) K 6 (16) replicas of the systems were taken to sample the imaginary time path integral, which has been shown to yield converged quantum kinetic energies [50,51]. Molecular dynamics trajectories were generally 10 ps long, which we show in the SI (section II) are sufficiently long to obtain converged binding free energies. For DFT we used the optB88-vdW functional,[52] which is a revised version of the van der Waals density functional (vdW-DF) of Dion *et al.*[53] This functional is particularly appropriate for the base pair duplexes under consideration as it yields very accurate interaction energies for the complexes in comparison to quantum chemistry reference calculations (see SI section II). In the SI (section III) we also report results from the hybrid PBE0 functional[54] for the binding free energy change of the AT base pair. Within the statistical error bars the results obtained with PBE0 and optB88-vdW are the same at room temperature.

The well-known structures of the Watson-Crick base pair complexes are shown in Fig. 1.

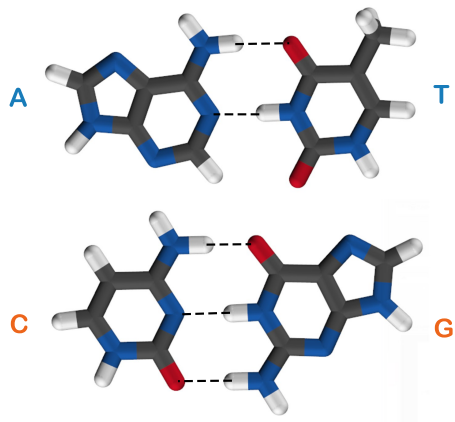

Figure 1: Structures of the Watson-Crick adenine-thymine (AT) and cytosine-guanine (CG) base pairs. Black: carbon; Red: oxygen; Blue: nitrogen; White: hydrogen.

From this it can be seen that the AT complex is held together by two H bonds (a NH-O and a NH-N bond), whereas the CG complex is held together by three (two NH-O bonds and a NH-N bond). The H bonds have a range of lengths, with both the NH-O and NH-N bonds varying from 1.7 to 1.9 Å in the ground state (geometry optimized) structure.

As a first step to understand the role of NQEs we ran a set of *ab initio* molecular dynamics (AIMD) simulations as well as a set of *ab initio* PIMD simulations. We concentrated on room temperature (300 K) and a cryogenic temperature (100 K) – room temperature is of obvious interest to biology and cryogenic temperatures are relevant to e.g. DNA preservation and DNA based devices.[56,57] At both temperatures the dimers remain hydrogen bonded and no intermolecular proton transfer is observed. Simply by comparing the structures obtained from the simulations with the classical and quantum nuclei we can gain an initial indication of the role NQEs play. Interestingly, we find that different HBs respond in a different manner to the inclusion of NQEs: some HBs get longer, some get shorter and some remain unchanged. Previously it was shown for a range of systems that how a HB responded to the inclusion of NQEs depended on its strength, with relatively strong H bonds becoming shorter and relatively weak HBs becoming longer.[18] In Fig. 2 we explore this issue for DNA base pairs by plotting how the intermolecular separation (specifically the N-O and N-N heavy atom distances) changes upon going from classical to quantum nuclei. HB strength is estimated with a simple and standard criterion involving the red shift in the harmonic stretching frequency of the covalent NH bond involved in the HB:[18,58,59] The larger the red shift of this stretching frequency, the stronger the HB. We find for the individual H bonds in the DNA complexes considered that the correlation seen before also holds here: the strong HBs do indeed tend to be shortened while the weak ones tend to be elongated by NQEs. Specifically, at 300 K, in the A-T base pair, the stronger NH-N bond becomes shorter in the PIMD simulations and the weaker NH-O bond barely changes. In the C-G base pair at 300 K, the weakest NH-O bond becomes longer in the PIMD simulations, while the other two H bonds become shorter. At 100 K, the correlation also holds and overall the H bond lengths change in a similar manner to what is observed at 300 K. It is clear, therefore, that NQEs impact on the intermolecular separation and this could be probed experimentally through e.g. isotopic substitution measurements of secondary geometric isotope effects. However from the structural data alone it remains unclear how NQEs have affected the interaction strength between the dimers in each complex.

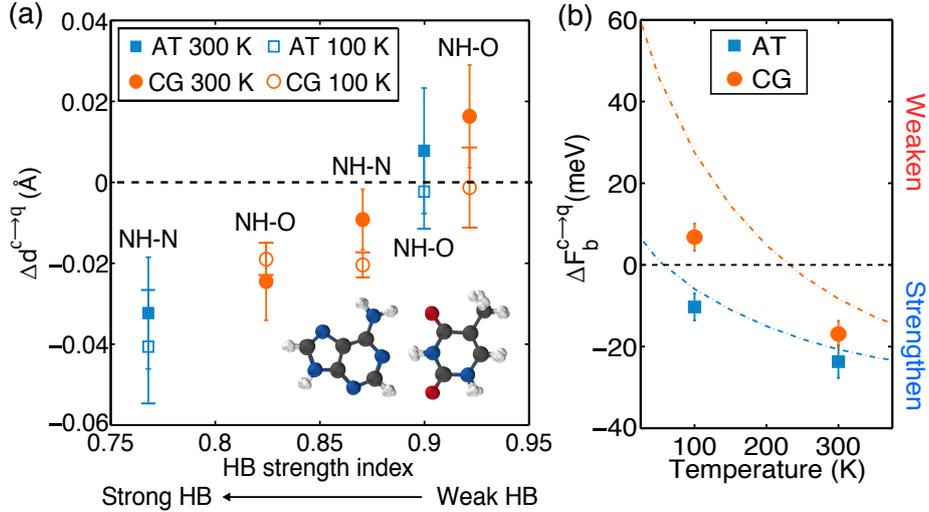

Figure 2: (a) Differences between the heavy atom separation distances from PIMD and MD simulations. Positive changes mean that the N(H)-O or N(H)-N bonds are longer in the PIMD than in the AIMD simulations; negative values that they are shorter in PIMD. The five different HBs in the base pairs are arranged from left to right in order of increasing strength; with strength characterized by the harmonic frequency of the N-H stretch in the HB divided by the harmonic frequency of the N-H stretch in the monomers.[18] A snapshot of the AT base pair taken from a PIMD simulation is also shown in the inset, each sphere is a "bead" in the PIMD simulation. (b) Plot of binding free energy change due to NQEs (Eq. 1) in the AT (blue) and CG (red) base pairs obtained from PIMD. A negative binding free energy change means NQEs strengthen the binding while a positive binding free energy change means NQEs weaken the binding. Also shown with the dashed lines are the predictions of each base pair obtained within the harmonic approximation. The error bars in (a) and (b) have been calculated using block averaging.[55]

In order to unambiguously determine how NQEs alter the interaction strength within the complexes we computed how the binding free energy changes upon going from classical to quantum nuclei. To this end we employed a thermodynamic integration scheme, previously used in calculations of isotope effects.[60–64] Full details of the specific approach used here are given in section I of the SI. The key point is that we obtain a binding free energy change by performing a thermodynamic integration in which the mass of the nuclei is switched from classical to quantum ($c \to$ q). Specifically the classical to quantum change in the binding free energy, $\Delta F_{\mathrm{b}}^{\mathrm{c \to q}}$, is given by

$$\Delta F_{\mathrm{b}}^{\mathrm{c \to q}} = 2 \int_0^1 \frac{\langle K_{\mathrm{M1:M2}}(g) \rangle - \langle K_{\mathrm{M1}}(g) \rangle - \langle K_{\mathrm{M2}}(g) \rangle}{g} \mathrm{d}g. \tag{1}$$

Here $\langle K \rangle$ is the ensemble average of the quantum kinetic energy, which can be directly obtained from PIMD simulations, and $g$ is a mass dependent integration variable. Separate PIMD simulations must be performed for the two isolated molecules, M1 and M2, and the hydrogen bonded complex M1:M2 and in order to obtain accurate values for the integrand simulations must be performed for several values of $g$ (7 in this study). In total to calculate $\Delta F_{\mathrm{b}}^{\mathrm{c \to q}}$ for a single system at a single temperature, trajectories equivalent to *ca.* 2 ns of *ab initio* MD simulations must be accumulated. This enormous computational cost is a key reason binding free energies have rarely been computed with *ab initio* PIMD.

The results obtained from the thermodynamic integration of our *ab initio* PIMD simulations are shown in Fig. 2 (b) at the two temperatures considered. A negative $\Delta F_{\mathrm{b}}^{\mathrm{c \to q}}$ means that NQEs strengthen the binding, while a positive value means NQEs weaken it. At room temperature the binding of both the AT and CG complexes is strengthened by $\sim$20 meV (0.5 kcal/mol) when NQEs are accounted for; specifically $24 \pm 4$ meV ($0.55 \pm 0.1$ kcal/mol) for AT and $17 \pm 4$ meV ($0.39 \pm 0.1$ kcal/mol) for CG. On an absolute scale of chemical bonding interactions these are, of course, small energies. However, as we know in biology energies tend to be finely balanced and very small changes in energy can be critical. For example, 20 meV is on the same scale as thermal energy at room temperature and similar to the cost to unzip double stranded DNA in solution (estimates range from 20 to 100 meV at room temperature).[36]

Upon considering how NQEs alter the energetics when the temperature decreases from 300 K to a cryogenic temperature, one would expect the influence to be greater than at room temperature. However, the opposite is the case with the contribution of NQEs to the binding being smaller at 100 K for both base pair complexes. In the AT complex NQEs strengthen the binding by $10 \pm 4$ meV ($0.23 \pm 0.1$ kcal/mol); about half the value at 300 K. In the CG complex the contribution of NQEs is even less ($7 \pm 4$ meV or $0.16 \pm 0.1$ kcal/mol) and, moreover, NQEs now act to destabilize the complex. Thus although simple assumptions about the temperature dependence of NQEs have been useful in understanding e.g. structural properties of liquid water[65] the same cannot be done when it comes to binding free energies.

Can we understand the free energy changes obtained? In the PIMD simulations the free energies arise from thermal sampling of the quantum kinetic energy through a complex interplay of vibrational modes, which is not particularly straightforward to interpret. It is possible to project the quantum kinetic energy on to particular modes (see e.g. refs. 25,30), however, here we opt to perform an analysis within the harmonic approximation, which provides a relatively straightforward means of establishing how specific groups of vibrational modes contribute to the observed changes in free energies. Within the harmonic approximation, the quantum kinetic energies in Eq. 1 are calculated from:

$$\langle K \rangle = \sum_i^{3N} \frac{\hbar \omega_i}{4} \coth \frac{\beta \hbar \omega_i}{2}, \tag{2}$$

where $\omega_i$ are the $3N$ harmonic vibrational frequencies (including the zero frequency translation and rotation modes) of the ground state geometry optimized complexes, $\hbar$ is the reduced Plank's constant, and $\beta$ is inverse temperature.[66] Results from the harmonic approximation are shown in Fig. 3. Clearly the harmonic approximation does not reproduce PIMD exactly; anharmonic effects are certainly important in these systems. Nonetheless, however, the harmonic approximation does reasonably well at a qualitative level: for both the AT and CG complexes the harmonic approximation gets the correct sign of the change and the correct temperature dependence. With the picture of competing quantum effects in mind we explored if it could also be used to explain the observed changes in binding free energies. To this end we define a separation between high frequency and low frequency vibrational modes; 2000 cm$^{-1}$ is chosen as this represents a threshold between high frequency covalent bond stretching modes and relatively low frequency bond bending and collective intermolecular stretching modes (see Fig. S3 in the SI). As shown on the right of each plot in Fig. 3 the high and low frequency vibrational modes have quite large ($ca.$ 40-90 meV) but opposing contributions to the overall binding free energy change: the low frequency modes reduce the binding free energy whereas the high frequency modes increase it. In the SI (section IV) we show the integration curves from which the histograms in Fig. 3 have been obtained, which show precisely how the observed changes in the quantum kinetic energy arise from competing contributions in the two vibrational regimes. In simple terms this behavior arises because the high frequency modes tend to be softened upon HB formation which reduces zero point energy, whereas the low frequency modes are hardened or new ones are created which tends to increase the zero point energy and so reduce the effective attraction. Overall it is clear that the net impact of NQEs on DNA binding results from a significant cancellation of two larger quantum contributions and that the picture of competing quantum effects can be used to quantitatively understand how NQEs alter HB strengths.

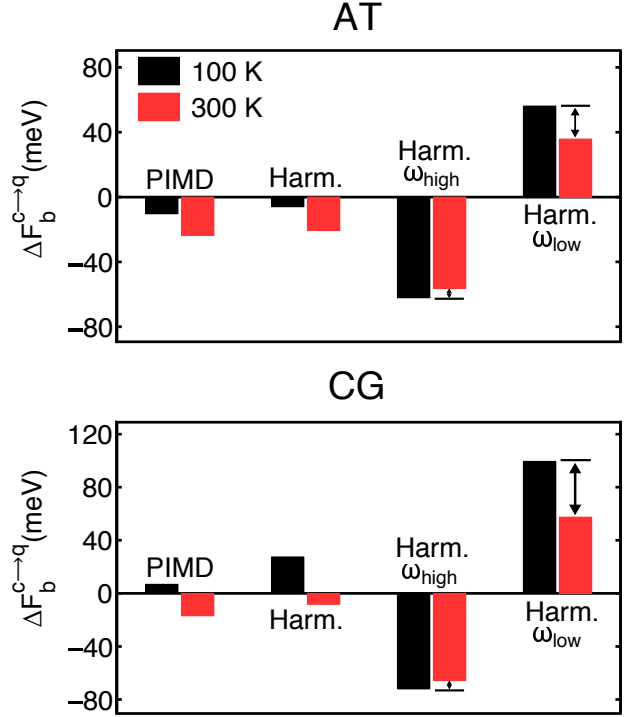

Figure 3: Competing quantum effects and explanation of the anomalous temperature dependence. The binding free energy change ($\Delta F_{\mathrm{b}}^{\mathrm{c} \rightarrow \mathrm{q}}$) obtained from the PIMD simulations is compared with results from the harmonic approximation (Harm.). The binding free energy changes within the harmonic approximation are also decomposed into high ($\omega_{\mathrm{high}}$) and low frequency ($\omega_{\mathrm{low}}$) contributions, revealing that the net change in binding free energy arises from a significant cancellation of contributions from these two regimes. The unusual temperature dependence simply arises because of a greater cancellation of terms at 100 K (black bars) than at 300 K (red bars). The change with temperature is more pronounced for the contribution from the low frequency modes than it is for the high frequency modes.

Let us now move on to the seemingly anomalous temperature dependence of NQEs on the binding free energies. Having established that the overall change in binding free energy arises from a cancellation of two opposing effects, one can recognise that there is a greater cancellation of terms at 100 K than there is at 300 K (Fig. 3). Looking at this figure more closely we can see that as the temperature increases from 100 K to 300 K, the contributions to the binding free energy differences from both the low frequency and high frequency modes decrease. This is to be expected and is consistent with conventional understanding that quantum effects are less prominent at high temperatures. However, the change with temperature is more pronounced for the low frequency modes than it is for the high frequency modes. These changes are governed by the occupation of the vibrational modes through the relation $k_{\mathrm{B}}T/\hbar\omega$, and within the temperature regime being considered the population of the high frequency modes changes much less than that of the low frequency modes. Hence it is the underlying competition coupled with the differing temperature dependence of the high and low frequency modes that makes the net impact of NQEs on the binding free energies more significant at 300 K than at 100 K. It is interesting to note that differences in the temperature dependence of the intermolecular and intramolecular modes have also been used to provide qualitatively the same explanation for isotope fractionation in water; in particular to explain an interesting inversion at high temperatures in the liquid water/vapor fractionation ratio.[25,30]

The temperature dependence of NQEs is of relevance beyond the HBs in DNA and we now show how a minimalistic model of HB formation can be used to make predictions about the temperature dependence of NQEs in HBs in general. In the model we mimic the essence of the competing quantum effects with only two variables: $\omega_{\mathrm{low}}$ and $\omega_{\mathrm{high}}$, which represent the total red shift of the high frequency modes and the total blue shift of the low frequency modes, respectively. (No explicit HB parameters i.e. bond length or model potentials are introduced here.) Upon using these two variables to compute the change in quantum kinetic energy (see Section V of the SI) one can predict whether NQEs strengthen or weaken the binding of a hydrogen bonded system at a given temperature. In Fig. 4 we show how the transition from NQEs strengthening HBs to NQEs weakening HBs depends on the interplay of these modes at room temperature (solid line) and at 100 K (dashed line); a broader range of temperatures is reported in the SI (section V). Also included in Fig. 4 are the results for some specific hydrogen bonded dimers in which all of the modes are taken into account. The particular dimers considered include the two DNA base pair complexes as well as water, HF, formamide, and formic acid dimers. The model presented here is incredibly simple, for example it neglects changes in modes at intermediate frequencies,[31] however, nonetheless, it produces qualitatively correct results. It correctly places the base pairs just within the strengthening regime at room temperature and (in agreement with full harmonic frequency calculations) that the water, HF, and formamide dimers are weakened at room temperature. It also correctly captures the behavior observed for the CG dimer wherein NQEs switch from strengthening to weakening upon lowering the temperature. As well as CG, the formic acid dimer also behaves in a similar manner, revealing that CG is not any sort of a special case and other hydrogen bonded systems could exhibit similar behavior. Although *Ab initio* PIMD simulations are becoming increasingly affordable computationally,[49] on the whole they remain expensive and are far from routine. However, the conceptual framework presented here allows for ballpark predictions to be made of the role of NQEs in HBs on the basis of (harmonic) vibrational frequencies. This is data that can be obtained from experiment or relatively cheap and easy vibrational frequency calculations.

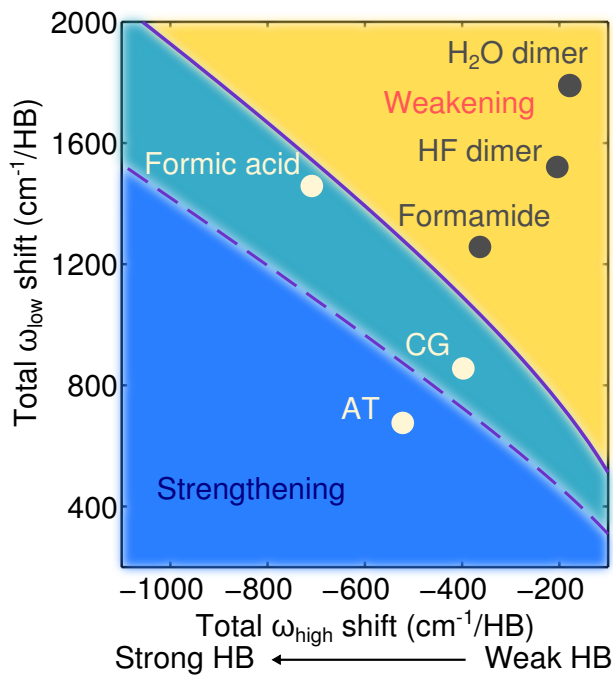

Figure 4: The temperature at which NQEs switch from weakening to strengthening the binding for the model hydrogen bonded system, plotted as a function of total high ($\omega_{\text{high}}$) and total low ($\omega_{\text{low}}$) frequency mode shifts. The solid line marks the 300 K transition, whereas the dashed line marks the 100 K transition. Actual frequency changes corresponding to six specific dimers are also indicated on the figure; these data points correspond to the average changes per HB for frequencies computed within the harmonic approximation. At room temperature the AT and CG base pairs and the formic acid dimer are in the strengthening regime and the water dimer, HF dimer and Formamide dimer are in the weakening regime.

Before concluding it is useful to put the results and model obtained in the current study into context. The model presented here focuses on energetics and so it complements other successful HB models, such as the recently proposed diabatic model of McKenzie and co-workers.[21–23,67] We note that these models have been very useful in predicting geometric isotopic effects, proton transport and frequency shifts and so together with the current model a broader picture of the structure, dynamics and energetics of hydrogen bonds is emerging. Here we have obtained free energy changes upon moving from the classical to quantum regime and have established the absolute contribution from nuclear quantum effects. This is not a measure that is readily accessible in experiments where indications about the importance of quantum nuclear effects are obtained indirectly through isotopic substitution measurements. With this in mind we note that we have also calculated the H/D isotope effect on the binding free energy in the harmonic approximation. From this we find that at room temperature the binding free energy of the AT base pair decreases by 3 meV upon deuteration and the binding of the CG base pair increases by 6 meV upon deuteration. Thus, as expected, the difference between H and D does not capture the full contribution of nuclear quantum effects. Finally, we note again that biological environments will be more complex than the gas phase systems considered here and that the presence of a solvent will now doubt have an impact on the nature of the HBs. For example, experiments have shown that the proton transfer rate for base pair complexes varies with the dielectric constant of the solvent.[5,68,69] Our model does not directly address measurements such as these but we do expect the physical insight obtained here to hold in more complex environments and, to first order, estimates of the influence of a solvent could be made by examining how the solvent alters the vibrational frequencies within the hydrogen bonded complexes being considered.

To conclude, we have reported what is to the best of our knowledge the first determinations of the quantum contribution to the binding free energy of DNA base pairs. We have found that NQEs strengthen the binding of both AT and CG complexes at room temperature. At a lower temperature, however, NQEs have a smaller impact on the binding free energies – analysis of the quantum kinetic energies in each system reveals that this seemingly anomalous temperature dependence arises from a balance between competing quantum effects associated with low frequency and high frequency modes of vibration. Upon forming a HB the high frequency (covalent bond stretching) modes are softened, hence quantum kinetic energy is lost and the system is stabilized. This stabilization, however, is offset by the quantum kinetic energy gained when low frequency modes are hardened or created upon forming the HB. This shows that the picture of competing quantum effects can be applied to understand how NQEs alter the energetics of hydrogen bonding and with this insight we have presented a simple model to estimate the temperature dependence of NQEs in hydrogen bonded systems in general. Of course real DNA is much more complex than the simple dimers considered here but, at the very least, the current study demonstrates that the role played by quantum effects could be more significant than previously anticipated and deserves further study.

# Acknowledgement

The authors would like to thank M. Ceriotti, T. E. Markland and P. Pedevilla for discussions and useful suggestions. J.C. and A.M. are supported by the European Research Council under the European Union's Seventh Framework Programme (FP/2007-2013) / ERC Grant Agreement number 616121 (HeteroIce project). A.M. is also supported by the Royal Society through a Royal Society Wolfson Research Merit Award. M.R. acknowledges funding from the German Research Foundation (DFG) under project RO 4637/1-1. Y.-X.F. and X.-Z.L. are supported by the National Science Foundation of China under Grant Nos 11275008, 11422431 and the China Postdoctoral Science Foundation under Grand No. 2014M550005. Via our membership of the UK's HEC Materials Chemistry Consortium, which is funded by EPSRC (EP/L000202), this work used the ARCHER UK National Supercomputing Service (http://www.archer.ac.uk).

## Supporting Information Available

The Supporting Information is available free of charge.

- Details on the mass-TI method; tests of the computational setup; results for the binding free energy calculation using the PBE0 functional; discussions on the thermodynamic integrand curves obtained from the mass-TI simulations; discussion of the model for predicting NQEs in HBs.

This material is available free of charge via the Internet at `http://pubs.acs.org/`.

## References

(1) Jeffrey, G. *An Introduction to Hydrogen Bonding*; Topics in Physical Chemistry - Oxford University Press; Oxford University Press, 1997.

(2) Latimer, W. M.; Rodebush, W. H. Polarity and Ionization from the Standpoint of the Lewis Theory of Valence. *J. Am. Chem. Soc.* **1920**, *42*, 1419–1433.

(3) Arunan, E.; Desiraju, G. R.; Klein, R. A.; Sadlej, J.; Scheiner, S.; Alkorta, I.; Clary, D. C.; Crabtree, R. H.; Dannenberg, J. J.; Hobza, P. et al. Definition of the Hydrogen Bond (IUPAC Recommendations 2011). *Pure Appl. Chem.* **2011**, *83*, 1637–1641.

(4) Douhal, A.; Kim, S. K.; Zewail, A. H. Femtosecond Molecular Dynamics of Tautomerization in Model Base Pairs. *Nature* **1995**, *378*, 260–263.

(5) Kwon, O.-H.; Zewail, A. H. Double Proton Transfer Dynamics of Model DNA Base Pairs in the Condensed Phase. *Proc. Natl. Acad. Sci. U.S.A.* **2007**, *104*, 8703–8708.

(6) Pérez, A.; Tuckerman, M. E.; Hjalmarson, H. P.; von Lilienfeld, O. A. Enol Tautomers of Watson-Crick Base Pair Models Are Metastable Because of Nuclear Quantum Effects. *J. Am. Chem. Soc.* **2010**, *132*, 11510–11515.

(7) Hwang, J.-K.; Warshel, A. How Important Are Quantum Mechanical Nuclear Motions in Enzyme Catalysis? *J. Am. Chem. Soc.* **1996**, *118*, 11745–11751.

(8) Billeter, S. R.; Webb, S. P.; Agarwal, P. K.; Iordanov, T.; ; Hammes-Schiffer, S. Hydride Transfer in Liver Alcohol Dehydrogenase: Quantum Dynamics, Kinetic Isotope Effects, and Role of Enzyme Motion. *J. Am. Chem. Soc.* **2001**, *123*, 11262–11272.

(9) Pu, J.; Gao, J.; ; Truhlar, D. G. Multidimensional Tunneling, Recrossing, and the Transmission Coefficient for Enzymatic Reactions. *Chem. Rev.* **2006**, *106*, 3140–3169.

(10) Major, D. T.; Heroux, A.; Orville, A. M.; Valley, M. P.; Fitzpatrick, P. F.; Gao, J. L. Differential Quantum Tunneling Contributions in Nitroalkane Oxidase Catalyzed and the Uncatalyzed Proton Transfer Reaction. *Proc. Natl. Acad. Sci. U.S.A.* **2009**, *106*, 20734–20739.

(11) Glowacki, D. R.; Harvey, J. N.; Mulholland, A. J. Taking Ockham's Razor to Enzyme Dynamics and Catalysis. *Nature Chem.* **2012**, *4*, 169–176.

(12) Wang, L.; Fried, S. D.; Boxer, S. G.; Markland, T. E. Quantum Delocalization of Protons in the Hydrogen-Bond Network of an Enzyme Active Site. *Proc. Natl. Acad. Sci. U.S.A.* **2014**, *111*, 18454–18459.

(13) Ubbelohde, A. R.; Gallagher, K. J. Acid-Base Effects in Hydrogen Bonds in Crystals. *Acta. Crystallogr.* **1955**, *8*, 71–83.

(14) Benedict, H.; Limbach, H.-H.; Wehlan, M.; Fehlhammer, W.-P.; Golubev, N. S.; Janoschek, R. Solid State N(15) NMR and Theoretical Studies of Primary and Secondary Geometric H/D Isotope Effects on Low-Barrier NHN-Hydrogen Bonds. *J. Am. Chem. Soc.* **1998**, *120*, 2939–2950.

(15) Petrenko, V.; Whitworth, R. *Physics of Ice*; Clarendon Press, 1999.

(16) McMahon, M.; Nelmes, R.; Kuhs, W.; Semmingsen, D. The Effect of Deuteration on the Crystal Structure of Squaric Acid (H2C4O4) in its Ordered Phase. *Z. Kristallogr.* **2010**, *195*, 231–239.

(17) Wikfeldt, K. T.; Michaelides, A. Communication: Ab Initio Simulations of Hydrogen-Bonded Ferroelectrics: Collective Tunneling and the Origin of Geometrical Isotope Effects. *J. Chem. Phys.* **2014**, *140*, 041103.

(18) Li, X. Z.; Walker, B.; Michaelides, A. Quantum Nature of the Hydrogen Bond. *Proc. Natl. Acad. Sci. U.S.A.* **2011**, *108*, 6369.

(19) Zeidler, A.; Salmon, P. S.; Fischer, H. E.; Neuefeind, J. C.; Simonson, J. M.; Lemmel, H.; Rauch, H.; Markland, T. E. Oxygen as a Site Specific Probe of the Structure of Water and Oxide Materials. *Phys. Rev. Lett.* **2011**, *107*, 145501.

(20) Zeidler, A.; Salmon, P. S.; Fischer, H. E.; Neuefeind, J. C.; Simonson, J. M.; Markland, T. E. Isotope Effects in Water as Investigated by Neutron Diffraction and Path Integral Molecular Dynamics. *J. Phys. Condens. Matter* **2012**, *24*, 284126.

(21) McKenzie, R. H. A Diabatic State Model for Donor-Hydrogen Vibrational Frequency Shifts in Hydrogen Bonded Complexes. *Chem. Phys. Lett.* **2012**, *535*, 196–200.

(22) McKenzie, R. H.; Bekker, C.; Athokpam, B.; Ramesh, S. G. Effect of Quantum Nuclear Motion on Hydrogen Bonding. *J. Chem. Phys.* **2014**, *140*, 174508.

(23) McKenzie, R. H.; Athokpam, B.; Ramesh, S. G. Isotopic Fractionation in Proteins as a Measure of Hydrogen Bond Length. *J. Chem. Phys.* **2015**, *143*, 044309.

(24) Habershon, S.; Markland, T. E.; Manolopoulos, D. E. Competing Quantum Effects in the Dynamics of a Flexible Water Model. *J. Chem. Phys.* **2009**, *131*, 024501.

(25) Markland, T. E.; Berne, B. J. Unraveling Quantum Mechanical Effects in Water Using Isotopic Fractionation. *Proc. Natl. Acad. Sci. U.S.A.* **2012**, *109*, 7988–7991.

(26) Alfè, D.; Gillan, M. J. Ab Initio Statistical Mechanics of Surface Adsorption and Desorption. II. Nuclear quantum effects. *J. Chem. Phys.* **2010**, *133*, 044103.

(27) Pamuk, B.; Soler, J. M.; Ramírez, R.; Herrero, C. P.; Stephens, P. W.; Allen, P. B.; Fernández-Serra, M.-V. Anomalous Nuclear Quantum Effects in Ice. *Phys. Rev. Lett.* **2012**, *108*, 193003.

(28) Liu, J.; Andino, R. S.; Miller, C. M.; Chen, X.; Wilkins, D. M.; Ceriotti, M.; Manolopoulos, D. E. A Surface-Specific Isotope Effect in Mixtures of Light and Heavy Water. *J. Phys. Chem. C* **2013**, *117*, 2944–2951.

(29) Ceriotti, M.; Cuny, J.; Parrinello, M.; Manolopoulos, D. E. Nuclear Quantum Effects and Hydrogen Bond Fluctuations in Water. *Proc. Natl. Acad. Sci. U.S.A.* **2013**, *110*, 15591–15596.

(30) Wang, L.; Ceriotti, M.; Markland, T. E. Quantum Fluctuations and Isotope Effects in Ab Initio Descriptions of Water. *J. Chem. Phys.* **2014**, *141*, 104502.

(31) Rossi, M.; Fang, W.; Michaelides, A. Stability of Complex Biomolecular Structures: van der Waals, Hydrogen Bond Cooperativity, and Nuclear Quantum Effects. *J. Phys. Chem. Lett.* **2015**, *6*, 4233–4238.

(32) Romanelli, G.; Ceriotti, M.; Manolopoulos, D. E.; Pantalei, C.; Senesi, R.; Andreani, C.

Direct Measurement of Competing Quantum Effects on the Kinetic Energy of Heavy Water upon Melting. *J. Chem. Phys. Lett.* **2013**, *4*, 3251–3256.

(33) Walker, B.; Michaelides, A. Direct Assessment of Quantum Nuclear Effects on Hydrogen Bond Strength by Constrained-Centroid Ab Initio Path Integral Molecular Dynamics. *J. Chem. Phys.* **2010**, *133*, 174306.

(34) Danilowicz, C.; Kafri, Y.; Conroy, R. S.; Coljee, V. W.; Weeks, J.; Prentiss, M. Measurement of the Phase Diagram of DNA Unzipping in the Temperature-Force Plane. *Phys. Rev. Lett.* **2004**, *93*, 078101.

(35) Danilowicz, C.; Coljee, V. W.; Bouzigues, C.; Lubensky, D. K.; Nelson, D. R.; Prentiss, M. DNA Unzipped Under a Constant Force Exhibits Multiple Metastable Intermediates. *Proc. Natl. Acad. Sci. U.S.A.* **2003**, *100*, 1694–1699.

(36) SantaLucia, J. A Unified View of Polymer, Dumbbell, and Oligonucleotide DNA Nearest-Neighbor Thermodynamics. *Proc. Natl. Acad. Sci. U.S.A.* **1998**, *95*, 1460.

(37) Wartell, R. M.; Benight, A. S. Thermal Denaturation of DNA Molecules: A Comparison of Theory with Experiment. *Phys. Rep.* **1985**, *126*, 67.

(38) Seeman, N. C. DNA in a Material World. *Nature* **2003**, *421*, 427–431.

(39) Peyrard, M.; Bishop, A. R. Statistical Mechanics of a Nonlinear Model for DNA Denaturation. *Phys. Rev. Lett.* **1989**, *62*, 2755–2758.

(40) Tuckerman, M. E.; Marx, D.; Klein, M. L.; Parrinello, M. On the Quantum Nature of the Shared Proton in Hydrogen Bonds. *Science* **1997**, *275*, 817–820.

(41) Benoit, M.; Marx, D.; Parrinello, M. Tunnelling and Zero-Point Motion in High-Pressure Ice. *Nature* **1998**, *392*, 258–261.

(42) Marx, D.; Tuckerman, M. E.; Hutter, J.; Parrinello, M. The Nature of the Hydrated Excess Proton in Water. *Nature* **1999**, *397*, 601–604.

(43) Geissler, P. L.; Dellago, C.; Chandler, D.; Hutter, J.; Parrinello, M. Autoionization in Liquid Water. *Science* **2001**, *291*, 2121–2124.

(44) Tuckerman, M. E.; Marx, D.; Parrinello, M. The Nature and Transport Mechanism of Hydrated Hydroxide Ions in Aqueous Solution. *Nature* **2002**, *417*, 925–929.

(45) Raugei, S.; Klein, M. L. Nuclear Quantum Effects and Hydrogen Bonding in Liquids. *J. Am. Chem. Soc.* **2003**, *125*, 8992–8993.

(46) Hassanali, A. A.; Cuny, J.; Ceriotti, M.; Pickard, C. J.; Parrinello, M. The Fuzzy Quantum Proton in the Hydrogen Chloride Hydrates. *J. Am. Chem. Soc.* **2012**, *134*, 8557–8569.

(47) VandeVondele, J.; Krack, M.; Mohamed, F.; Parrinello, M.; Chassaing, T.; Hutter, J. Quickstep: Fast and Accurate Density Functional Calculations Using a Mixed Gaussian and Plane Waves Approach. *Comput. Phys. Commun.* **2005**, *167*, 103–128.

(48) Hutter, J.; Iannuzzi, M.; Schiffmann, F.; VandeVondele, J. CP2K: Atomistic Simulations of Condensed Matter Systems. *Wiley Interdiscip. Rev. Comput. Mol. Sci.* **2014**, *4*, 15–25.

(49) Ceriotti, M.; More, J.; Manolopoulos, D. E. i-PI: A Python Interface for Ab Initio Path Integral Molecular Dynamics simulations. *Comput. Phys. Commun.* **2014**, *185*, 1019–1026.

(50) Ceriotti, M.; Manolopoulos, D. E. Efficient First-Principles Calculation of the Quantum Kinetic Energy and Momentum Distribution of Nuclei. *Phys. Rev. Lett.* **2012**, *109*, 100604.

(51) Ceriotti, M.; Manolopoulos, D. E.; Parrinello, M. Accelerating the Convergence of Path Integral Dynamics with a Generalized Langevin Equation. *J. Chem. Phys.* **2011**, *134*, 084104.

(52) Klimeš, J.; Bowler, D. R.; Michaelides, A. Chemical Accuracy for the van der Waals Density Functional. *J. Phys.: Condens. Matter* **2010**, *22*, 022201.

(53) Dion, M.; Rydberg, H.; Schröder, E.; Langreth, D. C.; Lundqvist, B. I. Van der Waals Density Functional for General Geometries. *Phys. Rev. Lett.* **2004**, *92*, 246401.

(54) Adamo, C.; Barone, V. Toward Reliable Density Functional Methods Without Adjustable Parameters: The PBE0 Model. *J. Chem. Phys.* **1999**, *110*, 6158–6170.

(55) Flyvbjerg, H.; Petersen, H. G. Error Estimates on Averages of Correlated Data. *J. Chem. Phys.* **1989**, *91*, 461–466.

(56) Hopkins, D. S.; Pekker, D.; Goldbart, P. M.; Bezryadin, A. Quantum Interference Device Made by DNA Templating of Superconducting Nanowires. *Science* **2005**, *308*, 1762–1765.

(57) Jones, M. R.; Seeman, N. C.; Mirkin, C. A. Programmable Materials and the Nature of the DNA Bond. *Science* **2015**, *347*, 1260901.

(58) Xantheas, S. S.; Dunning, T. H. Ab Initio Studies of Cyclic Water Clusters (H2O)n, n=1-6. I. Optimal Structures and Vibrational Spectra. *J. Chem. Phys.* **1993**, *99*, 8774–8792.

(59) Cubero, E.; Orozco, M.; Hobza, P.; Luque, F. J. Hydrogen Bond versus Anti-Hydrogen Bond: A Comparative Analysis Based on the Electron Density Topology. *J. Phys. Chem. A* **1999**, *103*, 6394–6401.

(60) Vaníček, J.; Miller, W. H. Efficient Estimators for Quantum Instanton Evaluation of the Kinetic Isotope Effects: Application to the Intramolecular Hydrogen Transfer in Pentadiene. *J. Chem. Phys.* **2007**, *127*, 114309.

(61) Pérez, A.; von Lilienfeld, O. A. Path Integral Computation of Quantum Free Energy

Differences Due to Alchemical Transformations Involving Mass and Potential. *J. Chem. Theory Comput.* **2011**, *7*, 2358–2369.

(62) Ceriotti, M.; Markland, T. E. Efficient Methods and Practical Guidelines for Simulating Isotope Effects. *J. Chem. Phys.* **2013**, *138*, 014112.

(63) Marsalek, O.; Chen, P.-Y.; Dupuis, R.; Benoit, M.; Méheut, M.; Bačić, Z.; Tuckerman, M. E. Efficient Calculation of Free Energy Differences Associated with Isotopic Substitution Using Path-Integral Molecular Dynamics. *J. Chem. Theory Comput.* **2014**, *10*, 1440–1453.

(64) Herrero, C. P.; Ramírez, R. Path-Integral Simulation of Solids. *J. Phys.: Condens. Matter* **2014**, *26*, 233201.

(65) Morrone, J. A.; Car, R. Nuclear Quantum Effects in Water. *Phys. Rev. Lett.* **2008**, *101*, 017801.

(66) Pavone, P.; Baroni, S. Dependence of the Crystal Lattice Constant on Isotopic Composition: Theory and Ab Initio Calculations for C, Si, and Ge. *Solid State Commun.* **1994**, *90*, 295–297.

(67) McKenzie, R. H. A Diabatic State Model for Double Proton Transfer in Hydrogen Bonded Complexes. *J. Chem. Phys.* **2014**, *141*, 104314.

(68) Takeuchi, S.; Tahara, T. The Answer to Concerted Versus Step-wise Controversy for the Double Proton Transfer Mechanism of 7-Azaindole Dimer in Solution. *Proc. Natl. Acad. Sci. U.S.A.* **2007**, *104*, 5285–5290.

(69) Nielsen, M. F.; Ingold, K. U. Kinetic Solvent Effects on Proton and Hydrogen Atom Transfers from Phenols. Similarities and Differences. *J. Am. Chem. Soc.* **2006**, *128*, 1172–1182.